\documentclass[aps,prl,twocolumn,showpacs,superscriptaddress]{revtex4}
\usepackage{graphicx}
\begin{document}

\title{Phonon Hall Effect in Nanoscale Four-Terminal Junctions}
     \author{Lifa~Zhang}
     \affiliation{Department of Physics and Centre for Computational Science and Engineering,
     National University of Singapore, Singapore 117542, Republic of Singapore }
     \author{Jian-Sheng~Wang}
     \affiliation{Department of Physics and Center for Computational Science and Engineering,
     National University of Singapore, Singapore 117542, Republic of Singapore }
     \author{Baowen~Li}
      \altaffiliation{Electronic address: phylibw@nus.edu.sg}
       \affiliation{Department of Physics and Center for Computational Science and Engineering,
     National University of Singapore, Singapore 117542, Republic of Singapore }
     \affiliation{NUS Graduate School for Integrative Sciences and Engineering,
      Singapore 117597, Republic of Singapore}
\date{9 May 2009}

\begin{abstract}
{Using an exact nonequilibrium Green's function formulism, the
phonon Hall effect (PHE) for paramagnetic dielectrics is studied in
a nanoscale four-terminal device setting. The temperature difference
in the transverse direction of the heat current is calculated for
two-dimensional models with the magnetic field perpendicular to the
plane.  We find that there is a PHE in nanoscale paramagnetic
dielectrics, the magnitude of which is comparable to the millimeter
scale experiments. If the dynamic matrix of the system satisfies
mirror reflection symmetry, the PHE disappears. The Hall temperature
difference changes sign if the magnetic field is sufficiently large
or if the size increases. }
\end{abstract}

\maketitle

In parallel to the study of electronics, phononics gets very active
and hot recently\cite{Wang2008}. Several conceptual phononic
(thermal) devices such as thermal diode, transistor, logic gate and
even thermal memory have been proposed to control phonon and process
information with phonons. The magnetic field is another degree of
freedom could be potentially used to control phonon transport
\cite{magnetothernal}.  Indeed, a novel phenomenon -- phonon Hall
effect (PHE) -- has been discovered experimentally by Strohm, et al.
\cite{Strohm2005}, which is an analog of the electrical Hall effect
for the heat flow in dielectrics.  The authors found a temperature
difference up to $200\,\mu$K between the sample edges in the
direction perpendicular to both the heat flow and the magnetic
field. This effect has been confirmed in \cite{Inyushkin2007}. The
electronic Hall effect is well understood in terms of Lorentz force.
However, since no charges are involved and phonons can not couple to
the magnetic field directly, it is thus not trial and straight
forward to understand the PHE. The magnetic field can polarize the
paramagnetic ions; the subsystem of isolated ions carrying magnetic
moment $M$ couples to phonons, this spin-phonon interaction (SPI)
determines the phonon Hall effect \cite{Sheng2006,Kagan2008}.

Theoretical models for PHE have been proposed in
Refs.~\cite{Sheng2006,Kagan2008}, in which phonons are treated
ballistically. However, according to Ref.\cite{Strohm2005}, the mean
free path ($1 \mu$m) is far less than the system size (15.7 mm),
therefore, it is not appropriate to treat the diffusive PHE with
ballistic theory. Moreover, in all the previous theoretical work,
the spin-phonon interactions were considered by perturbation theory,
and they are not consistent with each other.

In this Letter, we would like to address two questions: (1) whether
the PHE exists in nano-scale systems, or in other words whether the
diffusive PHE can be observed in the ballistic regime; (2)treat the
PHE by an exact - non-perturbative theory. The first question is in
fact not trivial at all. Since many physics laws valid in
macroscopic scale are not necessary true in nanoscale. For example,
Fourier law of heat conduction is broken down in
nanoscale\cite{FourierLaw}. As for the second question, we will take
the nonequilibrium Green's function (NEGF) approach. The NEGF is an
elegant and powerful method to treat nonequilibrium and interacting
systems in a rigorous way. NEGF is widely applied to the electronic
and thermal transport, and is successful to study the spin Hall
effect in junctions \cite{Sheng2005}.

\begin{figure}
\includegraphics[width=3.2 in,angle=0]{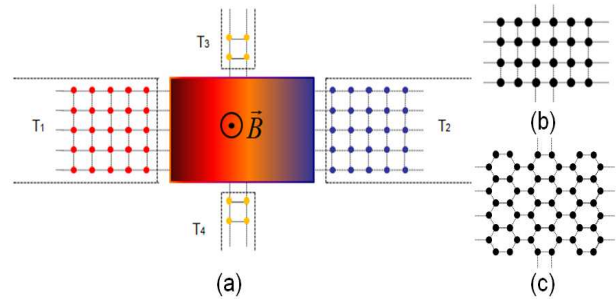}%
\caption{\label{fig1model}(color online) The four-terminal PHE setup
used for calculating the thermal conductance and the temperature
difference $T_3-T_4$.  (a)  The left and right leads have
temperatures $T_1$ and $T_2$, the upper and lower probe-leads have
temperatures $T_3$ and $T_4$. The center part can be different
lattices, such as square lattice (b) or honeycomb lattice (c).}
\end{figure}

To develop a nonperturbative theory for PHE in nanoscale four
terminal junctions, we consider a model as shown
Fig.~\ref{fig1model} by taking into account the actual measuring
process. The thermal conductance of the system can be calculated by
the NEGF method. As we shall see later, our model systems can
produce features similar to experiments, even though our systems are
of nanometer scale while the experimental systems are of millimeter
scale and are in the diffusive regime.

We consider the same Hamiltonian of SPI as in
Refs.~\cite{Sheng2006,Kagan2008}, which can be expressed in the form $ H_I =
g\sum\limits_n {\vec s_n \cdot (\vec U_n \times \vec P_n )}. $ Here, $\vec U_n
$ and $\vec P_n$ are the vectors of displacement and momentum of the $n$-th
lattice site.  In the presence of a magnetic field $\vec B$, each lattice site
has a magnetization $\vec M$. For isotropic SPI, the isospin $\vec s_n$ is
parallel to $\vec M_n$, and the ensemble average of the isospin is
proportional to the magnetization, that is $\langle \vec s_n \rangle = c \vec
M$.  Therefore, under the mean-field approximation, the SPI can be represented
as
\begin{equation}\label{hamspi}
H_I  = \sum\limits_n {\vec \Lambda  \cdot (\vec U_n  \times \vec
P_n )},
\end{equation}
where, $\vec \Lambda =gc\vec M$, has the units of frequency.  According to
\cite{Sheng2006}, $\Lambda$ is estimated to be $0.1\, {\rm cm}^{-1} \approx 3
\times 10^9 \,{\rm Hz} $ at $B = 1\, {\rm T}$ and $T = 5.45\,{\rm K}$, which
is within the possible range of the coupling strength in ionic insulators
\cite{Orbach1961,Manenkov1966}. In our calculation, we will use this relation
to map $\Lambda$ to magnetic field.

Now we turn to the derivation of the general formula for the
temperature difference in the system as illustrated in
Fig.~\ref{fig1model}, where a 2D lattice sample, which can be
honeycomb lattice or square lattice, is connected with four ideal
semi-infinite leads.  We denote the lattice as $N_R\times N_C$,
$N_R,\,N_C$ correspond to the number of rows and columns.  We adjust
the temperatures of upper and lower probes $T_3$ and $T_4$ such that
the heat currents from these two leads vanish, namely, $I_3=I_4=0$.
Then we get the relative Hall temperature difference as $R =
(T_3-T_4)/(T_1-T_2)$.

We treat the PHE problem by NEGF method \cite{negf} analogous to
those used in thermal transport \cite{refnegf}.  The total
Hamiltonian is assumed to be
\begin{equation}\label{hamh0}
H  = \sum\limits_{\alpha=0 }^4 H_{\alpha}
+ \sum\limits_{\beta=1}^4
{U_\beta ^T V_{\beta, 0} U_0 } +
  U_0^T AP_0,
\end{equation}
where $H_\alpha  = \frac{1}{2}\left(P_\alpha ^T P_\alpha + U_\alpha
^T K_\alpha  U_\alpha\right)$, and $A$ is an antisymmetric, block
diagonal matrix with the diagonal elements $ \left(
{\begin{array}{*{20}c}
   0 & {\Lambda  }  \\
   { - \Lambda  } & 0  \\
\end{array}} \right).$
Here, the integers 0 to 4 are associated with the center region,
left, right, upper, and lower leads, respectively.  $U_\alpha$ ($
P_\alpha$) are column vectors consisting of all the displacement
(momentum) variables in region $\alpha$. $K_\alpha$ is the spring
constant matrix and $V_{\beta, 0}=(V_{0, \beta})^T$ is the coupling
matrix between the $\beta$ lead and the central region.  The dynamic
matrix of the full linear system without SPI is
\begin{equation}\label{dynmatrix}
K = \left( {\begin{array}{*{20}c}
   {K_{1} } & 0 & {V_{1,0} } & 0 & 0  \\
   0 & {K_{3} } & {V_{3,0} } & 0 & 0  \\
   {V_{0,1} } & {V_{0,3} } & {K_0 } & {V_{0,4} } & {V_{0,2} }  \\
   0 & 0 & {V_{4,0} } & {K_{4} } & 0  \\
   0 & 0 & {V_{2,0} } & 0 & {K_{2} }  \\
\end{array}} \right).
\end{equation}
We obtain the equation for $U_0$ and $P_0$ as
\begin{eqnarray}
 \frac{{\partial U_0 (\tau )}}{{\partial \tau }} = P_0 (\tau ) - AU_0 (\tau
 ),\\
 \frac{{\partial P_0 (\tau )}}{{\partial \tau }} =  - K_0 U_0 (\tau )
 - \sum\limits_{\beta=1}^4  {V_{0,\beta } } U_\beta(\tau )   - AP_0 (\tau ).
 \end{eqnarray}
The energy flux to the central region from the lead $\alpha$ is,
\begin{equation}
I_\alpha   =  - \left\langle {\dot H_\alpha  } \right\rangle  =
\frac{i}{\hbar}\left\langle {[H_\alpha ,H] } \right\rangle ,\quad
\alpha=1,2,3,4.
\end{equation}
We define the contour-ordered Green's function as
$
 G^{\alpha \beta } (\tau ,\tau ') \equiv  - \frac{i}{\hbar}\left\langle
 {T_c\, U_\alpha  (\tau )U_\beta  (\tau ')^T } \right\rangle,
$ where $\alpha$ and $\beta$ refer to the region that the
coordinates belong to and $T_c$ is the contour-ordering operator.
Then the equations of motion of the contour ordered Green's function
can be derived.  In particular, the retarded Green's function for
the central region in frequency domain is
$
G^r [\omega ] = \Bigl[(\omega  + i\eta )^2  - K_0  - \Sigma^r [\omega ] -
A^2  + 2i\omega A\Bigr]^{ - 1}.
$
Here, $\Sigma^r=\sum\limits_{\alpha=1}^4 {\Sigma_\alpha ^r}$, and $
\Sigma_\alpha =V_{0, \alpha} g_\alpha V_{\alpha,0}$ is the self-energy due to
interaction with the heat bath, $ g_\alpha ^
r=[(\omega+i\eta)^2-K_\alpha]^{-1}$.  The lesser Green's function is obtained
through $G^< = G^r \Sigma^< G^a$ in the usual way.  We thus can calculate the
heat flux by the following formula,
\begin{equation}
I_\alpha  =  - \frac{1}{2\pi}\int_{ - \infty }^\infty \!\!d\omega\,
\hbar\omega  {\rm Re}\bigl[{\rm Tr}\bigl(G^r \Sigma _\alpha ^ <   +
G^ < \Sigma _\alpha ^a \bigr)\bigr].
\end{equation}
If the temperature differences among the leads are very small, we
can regard the system as at linear response regime, $T_\alpha = T +
\Delta _\alpha$. The linearized heat flux from each heat bath can be
written as
\begin{equation} \label{landauer-buttiker}
I_\alpha  = \sum\limits_{\beta=1}^4  {\sigma _{ \beta \alpha}
(\Delta _\alpha   - \Delta _\beta  )}.
\end{equation}
The conductance from heat bath $\alpha$ to $\beta$ is defined as
\begin{equation} \label{conductance}
\sigma _{ \beta \alpha } = \int_0^\infty\!\! {\frac{{d\omega }}{{2\pi
}}}\, \hbar\omega {T_{\beta \alpha } [\omega ]
{\frac{{\partial f}}{{\partial T}}}  },
\end{equation}
where, $T_{\beta \alpha } [\omega ] = {\rm{Tr}}(G^r \Gamma _\beta
G^a \Gamma _\alpha ),\;\;f = \bigl( e^{\hbar \omega / k_BT} -
1\bigr)^{-1}$, and $ \Gamma _\alpha = i\bigl(\Sigma _\alpha ^r
[\omega ] - \Sigma _\alpha ^a [\omega ]\bigr)$.  Therefore, if we
set the heat flux of lead 3 and lead 4 to zero, we can get the
relative Hall temperature difference as
\begin{eqnarray}
\!\!\!\!\!\!\!\!\!\! R &=& (\Delta _3  - \Delta _4 )/(\Delta _1
-  \Delta _2 )\nonumber \\
 &=&\!\! \frac{\sigma _{13} \sigma _{24}
  - \sigma _{23} \sigma _{14} }{(\sigma _{13}  + \sigma _{23}
   + \sigma _{43} )(\sigma _{14}  + \sigma _{24}
    + \sigma _{34} ) - \sigma _{43} \sigma _{34} }. \label{hdteq}
\end{eqnarray}
Equations (\ref{landauer-buttiker}) and (\ref{conductance}) are the
Landauer-B\"uttiker theory \cite{Datta,Buttiker} applied to the
multiple-lead thermal transport.

In the following calculation,  We assume a lattice constant
$a=2.465\,$\AA, and the force constant
$K_L=0.02394\,$eV/(amu$\cdot$\AA$^2$), $K_T=K_L/4$. The ratio of the
longitudinal and transverse sound speed to be $\delta=v_L/v_T\approx
\sqrt{K_L/K_T}=2$. Then the speed of sound for longitudinal acoustic
phonons is about $4000\,{\rm m/s}$. As mentioned above, $\Lambda$ is
estimated to be about $3 \times 10^9 \,{\rm Hz}\approx 2.0 \times
10^{-6}\, {\rm eV}$ at $B=1\,{\rm T}$.
\begin{figure}
\includegraphics[width=3.2 in,angle=0]{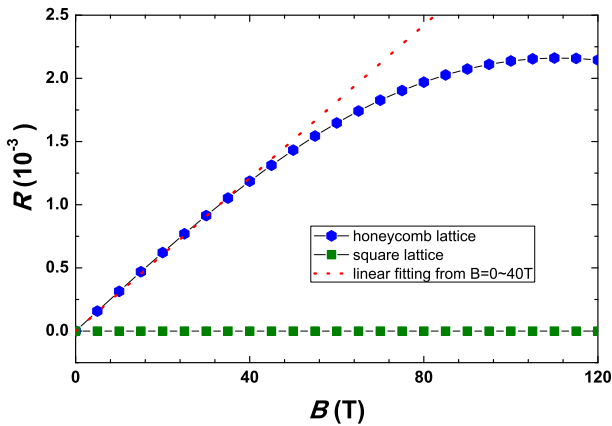}%
\caption{\label{fig2pheh}(color online) Hall temperature difference
$R$ versus magnetic field $B$ at temperature $T=5.45\,{\rm K}$. The
hexagon and square line correspond central regions for the honeycomb
and square lattices with a nearest-neighbor coupling. The red dotted
line corresponds to a linear fit from 0 to $40\,$T. The size of the
center region for honeycomb lattice is $9\times6$, the same with the
inset (c) in Fig.~\ref{fig1model}. }
\end{figure}

From Eq.~(\ref{hdteq}), we find that the relative Hall temperature
difference $R$ is an odd function of magnetic field.  Onsager
relation, $\sigma_{\alpha\beta}(\Lambda) =
\sigma_{\beta\alpha}(-\Lambda)$, always holds due to the definition
of the conductances. Furthermore, if there is a symmetry operation
$S$ such that
\begin{equation}\label{nophecnd}
 S\,K\,S^{ - 1}  = K,\quad S\,A\,S^{ - 1}  =  - A,
\end{equation}
then, $\sigma_{\alpha\beta}(\Lambda) = \sigma_{\alpha\beta}(-\Lambda)$.  If
this relation is true, then there is no phonon Hall effect in the system.
These symmetry considerations are consistent with a different treatment for
bulk systems based on Green-Kubo formula \cite{ourunpub}.

We discuss numerical results in the following.  Fig.~\ref{fig2pheh}
shows the temperature difference changing with magnetic field at
temperature $T=5.45\,{\rm K}$ for the honeycomb and square lattices
with nearest-neighbor couplings. For the honeycomb case, the Hall
temperature is odd and linear in the magnetic field between 0 and
$40\,$T, in that range the slope of the curve is $3\times10^{-5}\,
{\rm K/T}$, comparable to the experimental data in
Ref.~\cite{Strohm2005}. When the magnetic field is extremely large,
it will decrease. From our calculation, we find that the triangular
lattice has a similar behavior. However, for square lattice with the
nearest-neighbor coupling, there is no PHE at all. The spring
constant matrix between every nearest coupling sites is diagonal for
the square lattice. This matrix and also the full matrix $K$ is
invariant with respect to a reflection in $x$ or $y$ direction, thus
satisfying Eq.~(\ref{nophecnd}). If we consider next-neighbor
couplings of the lattice, the dynamic matrix $K$ will not have the
mirror reflection symmetry, and the PHE can come out.
\begin{figure}
\includegraphics[width=3.25 in,angle=0]{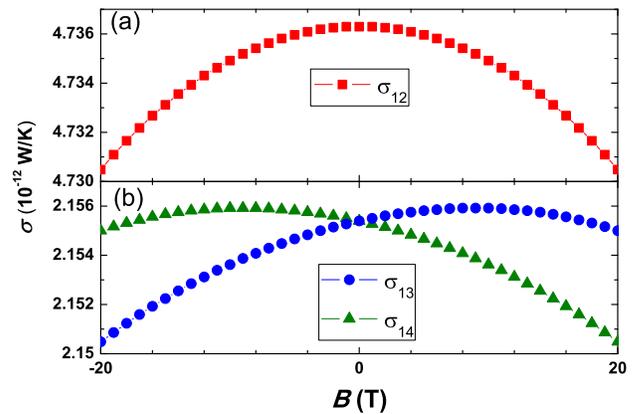}%
\caption{\label{fig3cond}(color online) Thermal
conductance versus the magnetic field at temperature $T=5.45{\rm K}$ for the
honeycomb lattice. (a) shows the conductance between two longitudinal leads
$\sigma _{12}$. (b) shows the conductance between one longitudinal lead and
one transverse probe-lead. The circle and triangular lines correspond to
$\sigma _{13}$ and $\sigma _{14}$, respectively. The size of center region is
$9\times6$. }
\end{figure}

We show the conductances among different leads in
Fig.~\ref{fig3cond}. We set all the couplings between the leads and
central region the same, and all the leads and central region have
the same spring constants for simplicity. Because of the symmetry of
the system, we have additional relations, $\sigma _{13} = \sigma
_{32} = \sigma _{24} = \sigma _{41} $, and $\sigma _{14} = \sigma
_{42} = \sigma _{23} = \sigma _{31} $.  We find that the conductance
between two longitudinal leads or two transverse probe-leads are
even in the magnetic field, which can be seen in
Fig.~\ref{fig3cond}(a), $\sigma_{34}$ has the same property.
However, for honeycomb lattice the conductance between one
longitudinal lead and one transverse probe-lead is not an even
function of magnetic field [Fig.~\ref{fig3cond}(b)], which gives
contribution to the Hall temperature difference.  Therefore, for
honeycomb lattice, the temperature difference is not zero.  But for
square lattice, $ \sigma _{13}$ is an even function of magnetic
field, the same is true for other components.  no PHE exists in such
systems.
\begin{figure}
\includegraphics[width=3.2 in,angle=0]{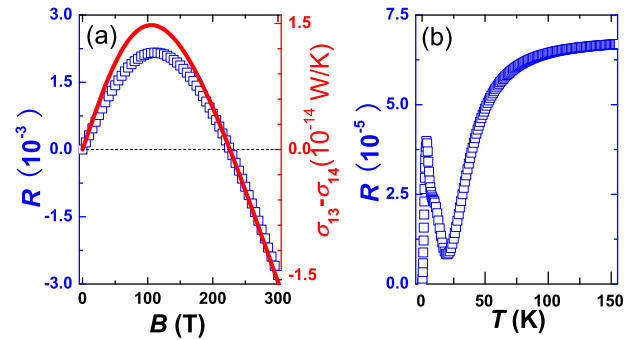}%
\caption{\label{fig4largehT}(color online) The relative  Hall
temperature difference $R$ versus the large magnetic field $B$ (a)
and high equilibrium temperature $T$ for honeycomb lattice (b). (a)
square shows $R$ changing with the magnetic field (left scale), the
red solid line shows the conductance difference
$\sigma_{13}-\sigma_{14}$ versus magnetic field (right scale). (b),
$R$ vs. equilibrium temperature at $B=1\,$T.}
\end{figure}

We show the numerical results for the ratio $R$  at $T=5.45{\rm K}$
for honeycomb lattice in Fig.~\ref{fig4largehT}(a). The temperature
difference will not be linear when the magnetic field is larger than
$40\,$T, after about $110\,$T, it will decrease, and about
$B_c\approx 230\, {\rm T}$, $R$ changes sign to negative.  It is the
same critical point for the difference of conductances
$\sigma_{13}-\sigma_{14}$, which is consistent with
Eq.~(\ref{hdteq}).   In Fig.~\ref{fig4largehT}(b), we show $R$
versus temperature at $B=1\, {\rm T}$. When the temperature
increases, $R$ will increase almost linearly. After some value, it
decreases, and then increases again. At last, it tends to a
constant. This behavior is due to the competition of the numerator
and denominator in Eq.~(\ref{hdteq}). When temperature is very high,
all the conductances tend to constants due to ballistic thermal
transport.

\begin{figure}
\includegraphics[width=3.2 in,angle=0]{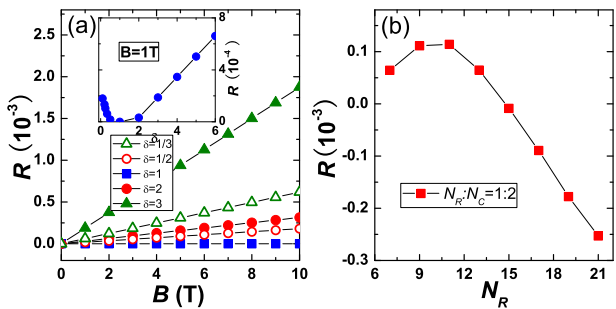}%
\caption{\label{fig5phed}(color online) (a) The relative  Hall
temperature difference versus magnetic field for different ratio of
the longitudinal and transverse sound speed $\delta=v_L/v_T$. The
inset shows $R$ versus $\delta$ at $B=1{\rm T}$. The data are
calculated for $9\times 6$ honeycomb lattice  at $T=5.45 {\rm K}$.
(b) The relative  Hall temperature difference versus the number of
rows of atoms for fixed aspect ratio $N_R:N_C=1:2$ at $B$=1 T and
$T=5.45$ K. }
\end{figure}

In Ref.~\cite{Sheng2006}, it is reported that $R$ decreases with
increasing  ratio of the longitudinal and transverse sound speed
$\delta=v_L/v_T$ and changes sign when $\delta$ becomes large than
5. However, we find that when the ratio ($\delta>1$) become large,
$R$ increases, see Fig.~\ref{fig5phed}(a). At exactly $\delta=1$,
when the longitudinal speed equals to the transverse speed, there is
no PHE, which testifies our condition, Eq.~(\ref{nophecnd}), for the
absence of PHE. All the spring constant matrices between the
nearest-neighbors become diagonal at $\delta=1$, the condition
Eq.~(\ref{nophecnd}) holds for a mirror reflection operation. If
$\delta<1$, $R$ increases again with the decreasing of $\delta$.
Although the ratio $R$ does not change sign with $\delta$, due to
the ballistic nature of a small system, the ratio $R$ is sensitive
to the geometric details, which is shown in Fig.~\ref{fig5phed}(b),
the magnitude and the sign of $R$ will change with the size
increasing.

In conclusion, a theory for PHE in nanoscale paramagnetic
dielectrics by NEGF approach is developed. The results are
consistent with the essential experimental features of PHE, such as
the magnitude and linear magnetic field dependence of the observed
transverse temperature difference. We find that there is no PHE if
the lattice satisfies a certain symmetry. The symmetry of the
dynamic matrix $K$ is the key point for the existence of PHE.  The
Hall temperature difference changes with equilibrium temperature and
tends to be a constant at last. And the Hall temperature difference
does not change sign with the ratio of the longitudinal and
transverse sound speed in the range of $\delta\in(0.1,10)$.

Most of our results should be verified by experiments on nano-scale
paramagnetic dielectrics, which can have potential applications to
controlling nanoscale phonon transport. For most paramagnetic
dielectric materials, because of the complexity of the coupling
(such as next or next-next neighboring interaction), the dynamic
matrix does not satisfy the mirror reflection symmetry, the PHE can
be present. The Hall temperature difference behavior with magnetic
field can be measured in a strong magnetic field (if $\Lambda=2.0
\times 10^{-5}$ eV, the magnetic field will be ten times smaller).
From our study, the PHE can be measured in nanoscale system at
relative high temperature ($\sim100$ K).

We thank Jinwu Jiang, Yonghong Yan, Jie Chen and Jie Ren for
fruitful discussions. L. Z. and B. L. are supported by the grant
R-144-000-203-112 from Ministry of Education of Republic of Singapore. J.-S. W. acknowledges support from a faculty
research grant R-144-000-173-112/101 of NUS.

\end{document}